\documentclass[hyper]{JHEP} 

\usepackage{epsfig}

\newcommand\fverb{\setbox\pippobox=\hbox\bgroup\verb}

\newcommand\fverbdo{\egroup\medskip\noindent%

            \fbox{\unhbox\pippobox}\ }

\newcommand\fverbit{\egroup\item[\fbox{\unhbox\pippobox}]}

\newbox\pippobox

\title{Hamiltonian Analysis of
    Non-Relativistic  Non-BPS Dp-brane}
\author{J. Kluso\v{n}\\
Department of
Theoretical Physics and Astrophysics\\
Faculty of Science, Masaryk University\\
Kotl\'{a}\v{r}sk\'{a} 2, 611 37, Brno\\
Czech Republic\\
E-mail: \email{klu@physics.muni.cz}} \preprint{}

 \abstract{We perform  Hamiltonian analysis of non-relativistic
    non-BPS Dp-brane. We find the constraint structure of this theory and  determine  corresponding
    equations of motion. We further discuss property of this theory at the
    tachyon vacuum.}

\newcommand{\bC}{\mathbf{C}}

\def\bA{\mathbf{A}}

\def\ttau{\tilde{\tau}}

\def\tF{\tilde{F}}
\def\bn{\mathbf{n}}
\def\tT{\tilde{T}}

\def\bB{\mathbf{B}}

\def\tx{\tilde{x}}

\def\be{\begin{equation}}

\def\ee{\end{equation}}

\def\bea{\begin{eqnarray}}

\def\eea{\end{eqnarray}}

\def\tmH{\tilde{\mH}}

\def\mH{\mathcal{H}}

\def\tG{\tilde{G}}

\def\tA{\tilde{A}}

\newcommand{\tZ}{\tilde{Z}}

\newcommand{\mK}{\mathcal{K}}

\newcommand{\mG}{\mathcal{G}}

\def \bA{\mathbf{A}}

\newcommand{\bT}{\mathbf{T}}

\newcommand{\ba}{\mathbf{a}}

\newcommand{\mL}{\mathcal{L}}

\def \tZ{\tilde{Z}}

\def\pb #1{\left\{#1\right\}}

\begin{document}
\section{Introduction and Summary}
Holography is very useful tool for the analysis of relativistic
strongly coupled field theory using Einstein classical
gravity in the bulk \cite{Maldacena:1997re,Gubser:1998bc,Witten:1998qj}.
One such an interesting application of the holography is the analysis of condensed matter systems, for recent review, see \cite{Hartnoll:2016apf}. It turns out that non-relativistic systems play fundamental role in this analysis and hence they are now studied very intensively since they are also related to famous P. Ho\v{r}ava's proposal
\cite{Horava:2009uw}, for recent review, see \cite{Wang:2017brl}
\footnote{See, for example
\cite{Bergshoeff:2017btm,Bergshoeff:2016lwr,Hartong:2015xda,
Bergshoeff:2015uaa,Andringa:2010it,Bergshoeff:2015ija,
Bergshoeff:2014uea}.}.

It is well known that non-relativistic symmetry can be imposed at the level of action for relativistic string or p-branes
\cite{Gomis:2000bd,Danielsson:2000gi,Gomis:2004pw,Gomis:2005pg,Kluson:2006xi,Batlle:2016iel,Gomis:2016zur}.
It is important to stress that different non-relativistic limits can be imposed, at least in principle. For example, "stringy" non-relativistic limit was firstly introduced in \cite{Gomis:2000bd, Danielsson:2000gi} where the time and one spatial directions along the string are large.

There is also another non-relativistic limit of the relativistic string where only the time direction is large
\cite{Batlle:2016iel,Gomis:2016zur}. It turns out that this corresponds to the situation where non-relativistic string does not vibrate. In fact, it represents a collection of non-relativistic massless particles with an energy density that depends on the position of the particle along the string \cite{Batlle:2016iel,Gomis:2016zur}. The action is invariant under Galilean transformations where the Poisson brackets of the generators of algebra close without central extension.

These different non-relativistic limits are very interesting
and could lead to new solvable sectors of string theory, as for example in \cite{Gomis:2005pg}. It is also interesting to study
how different string theory objects behave when they world-volume
actions are restricted by non-relativistic  limiting procedure. Certainly such an interesting object is an action for non-BPS Dp-brane
\cite{Sen:1999md,Bergshoeff:2000dq,Kluson:2000iy}. The question is whether different limiting procedures  of this action  do not spoil
remarkable properties of the tachyon effective  action. One such a nice property is that this action at its Hamiltonian formulation can describe the fate of non-BPS Dp-brane at its tachyon vacuum. For example, it was shown  that at the tachyon vacuum with constant electric flux Dp-brane
disappears
 and we are left with the gas of fundamental strings that move in the whole target space-time
\cite{Sen:2000kd,Sen:2003bc,Kluson:2016tgp} which agree with the general principles of tachyon condensation in open string theory
\footnote{It was shown in \cite{Kluson:2005dr} that in case of zero electric flux non-BPS Dp-brane at the tachyon vacuum  corresponds to the  gas of massless particles.}. The question is whether non-relativistic non-BPS Dp-brane action as was derived in
\cite{Kluson:2006xi} shares the same properties as its relativistic
precursor. It is possible that limiting procedure can change property of
non-BPS Dp-brane action. One such an example was studied recently in
\cite{Kluson:2017fam} where the Carrol limit of non-BPS Dp-brane was analyzed.
\footnote{Carroll limit of extended objects in string theory was
studied previously in \cite{Bergshoeff:2014jla,Bergshoeff:2015wma,Cardona:2016ytk}.}.
We showed there that the equations of motion of Carroll non-BPS Dp-brane at its tachyon vacuum have solutions that can be interpreted as fundamental strings with however important restrictions on their momenta so that there is not exact equivalence between  the conjecture that at the tachyon vacuum an
unstable Dp-brane disappears and we are left with the gas of Carroll strings moving in the whole target space-time.  As we will see in this paper in case of a non-relativistic non-BPS Dp-brane the situation is even worse. More explicitly, while the non-relativistic non-BPS Dp-brane action, as was derived in
\cite{Kluson:2006xi}, possesses tachyon kink solution that can be interpreted as stable D(p-1)-brane \cite{Kluson:2006xi},  in case
of its tachyon vacuum solution the situation is not the same. In order to show this we have to firstly determine Hamiltonian form of non-relativistic non-BPS action which is non-trivial task due to its complicated structure. We find that the bare Hamiltonian is proportional to Gauss constraint while the theory possesses $(p+1)-$primary   constraints. We calculate the algebra of these constraints and
show that they are the first class constraints which is a reflection of the fact that the action is invariant under world-volume diffeomorphism. Then we analyze equations of motion at the tachyon vacuum. We find, due to the structure of  Hamiltonian constraint, that corresponding solutions cannot be interpreted as the solutions describing the gas of non-relativistic strings. To see this explicitly we preform also Hamiltonian analysis of non-relativistic string in slightly different way from the original works  \cite{Gomis:2004ht,Batlle:2016iel}
that however coincide with them  when we introduce light-cone variables as in \cite{Batlle:2016iel}.

Let us outline our results derived in this paper. We perform Hamiltonian analysis of non-relativistic non-BPS Dp-brane action, determine the structure of the constraints. We derive corresponding equations of motion and we study their behavior at the tachyon vacuum. We show  that the equations of motion
of the non-relativistic non-BPS Dp-brane cannot be related to the equations of motion for non-relativistic string which can be a consequence of the  remarkable efficiency of the Dirac-Born-Infeld form of non-BPS Dp-brane action
\cite{Erkal:2009xq,Kutasov:2004ct,Kutasov:2003er,Sen:2007cz,Sen:2004nf}
that is lost when we perform non-relativistic contraction.

This paper is organized as follows. In the next section (\ref{second}) we review how to implement consistent non-relativistic limit for non-BPS Dp-brane. In section (\ref{third}) we perform Hamiltonian analysis of this theory and in section (\ref{fourth}) we determine algebra of constraints. In section (\ref{fifth}) we analyze corresponding equations of motion. In section (\ref{sixth}) we review non-relativistic limit in case of fundamental string and perform its Hamiltonian analysis. Finally in Appendix (\ref{A})
we review Hamiltonian analysis of non-relativistic string expressed using
light-cone variables.

\section{Non-Relativistic Limit of Non-BPS Dp-Brane}\label{second}
Our goal is to perform well defined  non-relativistic limit which is
consistent in the sense that all divergences that arises during the
limiting procedure, either cancel or give trivial contribution to
the action. It turns out that this procedure is rather non-trivial
and deserves careful treatment.  To see this in more details let us
consider a simple example when we start with a p-brane action in the
form
\begin{eqnarray}\label{actnongauge}
S&=&-\ttau_p\int d^{p+1}\xi
\sqrt{-\det \bA}
\ ,  \quad
\bA_{\alpha\beta}=\eta_{MN}\partial_\alpha \tx^M\partial_\beta \tx^N \   \nonumber \\
\end{eqnarray}
and try to implement non-relativistic limit, following recent paper
\cite{Batlle:2016iel}. It was argued there
that we can have $p+1$ different non-relativistic limits
according to the number of embedding coordinates
$(0,\dots,p)$ that are rescaled.  Explicitly, we have
\begin{eqnarray}
\tx^\mu=\omega X^\mu \ , \quad  \mu=0,\dots,q \ ,
\quad
\tx^i=X^i \ , \quad  i=q+1,\dots,d-1 \ ,\quad
\ttau=\frac{\tau}{\omega} \ ,
\nonumber \\
\end{eqnarray}
where we take $\omega\rightarrow \infty$ in the end. Note that the
matrix $\bA_{\alpha\beta}$ has the form
\begin{eqnarray}
\bA_{\alpha\beta}&=&\omega^2\tG_{\alpha\beta}+\ba_{\alpha\beta}  \ , \quad
\nonumber \\
\tG_{\alpha\beta}&= &\partial_\alpha X^\mu \partial_\beta X_\mu \ ,  \quad
\ba_{\alpha\beta}=\partial_\alpha X^i\partial_\beta X_i \ .  \nonumber \\
\end{eqnarray}
Now the problem is how to deal with this case since the matrix
 $\tG_{\alpha\beta}$ is singular
for $p\neq q$. This follows from the fact that it can be written as
\begin{equation}
\tG_{\alpha\beta}=E_\alpha^{ \mu}\eta_{\mu\nu}E^\nu_\beta  \ ,
\end{equation}
where $E_\alpha^\mu=\partial_\alpha X^\mu$ is $(p+1)\times (q+1)$ matrix that
has rank $(q+1)$. Since $\eta_{\mu\nu}$ is $(q+1)\times (q+1)$ matrix we find that the rank of the  matrix $\tG_{\alpha\beta}$  is $q+1$.  On the other hand since  $\tG_{\alpha\beta}$ is $(p+1)\times (p+1)$ matrix  we immediately find, since $q<p$, that $\tG_{\alpha\beta}$ is singular matrix so that it is not possible to introduce its inverse.  For that reason it is not clear how to introduce non-relativistic limit consistently for $q\neq p$ with the exception of the particle like non-relativistic limit that was analyzed in
\cite{Batlle:2016iel}.  Further, as was argued previously in
\cite{Gomis:2004pw,Gomis:2005pg}, we have to couple Dp-brane with a constant  Ramond-Ramond background in order to consistently remove divergences. Then in order to find well defined non-relativistic limit in case of non-BPS Dp-brane we have to follow an  analysis  performed in
\cite{Kluson:2006xi}.
To begin with we write Dirac-Born Infeld (DBI) part of  non-BPS Dp-brane action
in the flat background
\begin{eqnarray}\label{actnongauge1}
S_{DBI}&=&-\ttau_p\int d^{p+1}\xi
V(\tT)\sqrt{-\det \bA_{\alpha\beta}}
\ ,  \nonumber \\
& &\bA_{\alpha\beta}=\eta_{MN}\partial_\alpha \tx^M\partial_\beta \tx^N+
\tF_{\alpha\beta}+\partial_\alpha \tT\partial_\beta \tT \ , \quad  \tF_{\alpha\beta}=
\partial_\alpha \tA_\beta-\partial_\beta \tA_\alpha \ ,  \nonumber \\
\end{eqnarray}
where $\tx^M, M=0,\dots,d-1$ are embedding coordinates,  $\tA_\alpha,\alpha=0,\dots,p$ are world-volume gauge field, $\tT$ is the tachyon
with the potential $V(\tT)$ with the property that it is even function of $\tT$ that is zero for $\tT=\tT_{min}$ and $V(\tT=0)=1$. Let us now consider following
non-relativistic limit \cite{Gomis:2004pw,Kluson:2006xi}
\begin{eqnarray}\label{scaldef}
& & \tx^\mu=X^\mu \ , \quad \mu,\nu=0,\dots,p-1 \ , \nonumber \\
& &\tx^a=\lambda X^a \ , \quad  a=p,\dots, d-1 \ ,
\nonumber \\
& &\ttau_p=\lambda^{-2}\tau_p \ , \quad  \tF_{\alpha\beta}=\lambda F_{\alpha\beta} \ , \quad
\tA_\alpha=\lambda A_\alpha \ , \quad  \tT=T  \nonumber \\
\end{eqnarray}
and consider the limit $\lambda\rightarrow 0$ \footnote{Note an
important difference with the non-relativistic limit of stable
Dp-brane in the number of the embedding fields that are not scaled.
In case of non-BPS Dp-brane we have $p-$ these fields while in case
of stable Dp-brane the number of these fields is $p+1$. The reason
why one scalar mode is missing in case of non-BPS Dp-brane is  in
the specific form of the Wess-Zummino term for unstable Dp-brane
which is needed for the cancelation of the divergent term that
emerges
    when we perform  the non-relativistic limit on DBI  action. }.
 In other words we define relativistic limit as the limit when transverse fluctuations are small. Using (\ref{scaldef}) we find
 that the matrix $\bA_{\alpha\beta}$ has the form
\begin{equation}
\bA_{\alpha\beta}=\bB_{\alpha\beta}+\lambda F_{\alpha\beta}
 +\lambda^2 \bC_{\alpha\beta} \ ,
\end{equation}
where
\begin{eqnarray}
& &\bB_{\alpha\beta}=
\eta_{\mu\nu}\partial_\alpha X^\mu \partial_\beta X^\nu+
\partial_\alpha T\partial_\beta T \ , \quad
 \bC_{\alpha\beta}=\partial_\alpha X^a\partial_\beta X_a \ .
\nonumber \\
\end{eqnarray}
Then the DBI action has the form
\begin{eqnarray}\label{SDBIlimit}
S_{DBI}
&=&-\frac{\tau_p}{\lambda^2}
\int d^{p+1}\xi V(T)\sqrt{-\det \bB}-\nonumber \\
&-&\frac{\tau_p}{2}\int d^{p+1}\xi V(T)
\sqrt{-\det \bB}[\bB^{\alpha\beta}\partial_\beta X^a
\partial_\alpha X_a
-\frac{1}{2}\bB^{\alpha\beta}F_{\beta\gamma}
\bB^{\gamma\delta}F_{\delta \alpha}+O(\lambda^2)] \ ,
\nonumber \\
\end{eqnarray}
where we defined $\bB^{\alpha\beta}$ as inverse to $\bB_{\alpha\beta}$
\begin{equation}
\bB_{\alpha\beta}\bB^{\beta\gamma}=\delta^\gamma_\alpha \ .
\end{equation}
In order to  see that this is possible it is useful to   introduce following
 notation
\begin{equation}
Y^I\equiv (X^\mu,T)\ , \quad  Y^p=T \ , \quad  I,J=0,\dots,p
\end{equation}
so that  we can write $\bB_{\alpha\beta}$ as
\begin{equation}\label{defB}
\bB_{\alpha\beta}=\partial_\alpha Y^I\partial_\beta Y_I=E_\alpha^I
E_\beta^J\eta_{IJ} \ , \quad  I,J=0,\dots,p \ , \quad  \eta_{pp}=1 \ .
\end{equation}
Clearly $E_\alpha^I$ and $\eta_{IJ}$ are  $(p+1)\times(p+1)$ matrices so that  $\bB_{\alpha\beta}$ is non-singular matrix.

Using the notation introduced in (\ref{defB}) we can write the expression
on the first line in  (\ref{SDBIlimit}) as
\begin{eqnarray}
& &d^{p+1}\xi\sqrt{-\det \bB}
=d^{p+1}\xi\sqrt{-\det (E_\alpha^I\eta_{IJ}
E^J_\beta)}
\nonumber \\
& &=d^{p+1}\xi\det E_\alpha^I=E^0\wedge E^1\wedge \dots \wedge E^p=
-\frac{1}{(p+1)!}\epsilon_{I_0 \dots I_p}E^{I_0}\wedge
\dots E^{I_p} \ , \nonumber \\
\nonumber \\
\end{eqnarray}
where we used following normalization for Levi-Chivita symbol
\begin{equation}
\epsilon^{012\dots p}= 1 \ , \quad \epsilon_{012\dots p}=-1 \ ,
\end{equation}
and where $E^I=E^I_\alpha d\xi^\alpha $. In other words the divergent term is equal to
\begin{equation}\label{DBIdiv}
S_{DBI}^{div}=\frac{\tau_p^2}{\lambda^2(p+1)!}
\int V(T) \epsilon_{I_0\dots I_p}E^{I_0}\wedge \dots
\wedge E^{I_p} \ .
\end{equation}
Let us now consider Wess-Zummino term for unstable Dp-brane that expresses
the coupling of this brane to the background Ramond-Ramond forms
\begin{equation}
S_{WZ}=\ttau_p\int V(\tT)d\tT\wedge C \wedge e^{\tF} \ .
\end{equation}
Let us presume that there is background Ramond-Ramond $C_p$ form equal to
\begin{equation}
C_{\mu_0\dots\mu_{p-1}}=-(-1)^p\epsilon_{\mu_0\dots\mu_{p-1}} \ .
\end{equation}
For this background the WZ term has the form
\begin{eqnarray}
S_{WZ}& &
=-\frac{\tau_p}{\lambda^2}\int V(T)\frac{1}{p!}
\epsilon_{\mu_0\dots\mu_{p-1}}dX^{\mu_0}\wedge \dots
\wedge dX^{\mu_{(p-1)}}\wedge dT=\nonumber \\
&=&-\frac{\tau_p}{\lambda^2 (p+1)!}\int V(T)\epsilon_{I_0\dots I_p}
E^{I_0}\wedge \dots \wedge E^{I_p} \  \nonumber \\
\end{eqnarray}
and we see that it precisely cancels the divergent term (\ref{DBIdiv}). As a result we find well defined limited procedure that leads to   non-relativistic non-BPS Dp-brane action in  the form
\begin{eqnarray}\label{NRaction}
S_{NR}=
-\frac{\tau_p}{2}\int d^{p+1}\xi V(T)
\sqrt{-\det \bB}[\bB^{\alpha\beta}\partial_\beta X^a
\partial_\alpha X_a
-\frac{1}{2}\bB^{\alpha\beta}F_{\beta\gamma}
\bB^{\gamma\delta}F_{\delta \alpha}] \ .
\nonumber \\
\end{eqnarray}
This action was derived previously in \cite{Kluson:2006xi}. It was also shown
there that the tachyon kink solution of the equations of motion correctly
reproduces non-relativistic D(p-1)-brane action. Next step in the analysis of this action is to study its properties at the tachyon vacuum. In order to do this
we have to find
 Hamiltonian form of this action.
\section{Hamiltonian Formalism }\label{third}
The action (\ref{NRaction}) is rather complicated due to the fact
that  elements of the matrix $\bB_{\alpha\beta}$ depend on the time
derivative of $Y^I$. It turns out that it is useful to introduce
$(p+1)-$like decomposition of the matrix $\bB_{\alpha\beta}$ in
complete analogy with $p+1$ formalism in General Relativity
\cite{Gourgoulhon:2007ue,Arnowitt:1962hi} so that
\begin{eqnarray}
\bB^{\alpha\beta}&=&\left(\begin{array}{cc}\
-\frac{1}{N^2} & \frac{N^j}{N^2} \\
\frac{N^i}{N^2} & h^{ij}-\frac{N^iN^j}{N^2} \\ \end{array}
\right) \ , \nonumber \\
\bB_{\alpha\beta}&=&\left(\begin{array}{cc}
-N^2+N_ih^{ij}N_j & N_j \\
N_i & h_{ij} \\
\end{array}\right) \ .
\nonumber \\
\end{eqnarray}
If we compare $\bB_{\alpha\beta}$ defined above
with  $\bB_{\alpha\beta}=\partial_\alpha Y^I\partial_\beta Y_I$
we obtain an explicit relation between $N,N_i,h_{ij}$ and
derivatives of $Y^I$
\begin{eqnarray}
N_i&=&\partial_i Y^I\partial_0Y_I \ , \quad  h_{ij}=
\partial_i Y^I\partial_j Y_I \ , \nonumber \\
N^2&=&\partial_0 Y^I(\partial_i Y_I h^{ij}\partial_jY_J-
\delta_{IJ})\partial_0Y^J \equiv -\partial_0 Y^I V_{IJ}
\partial_0 Y^J \ , \nonumber \\
\end{eqnarray}
where $h^{ij}$ is inverse to $h_{ij}$ so that $h_{ij}h^{jk}=\delta_i^k$.
If we also introduce $(p+1)-$decomposition of the gauge field we obtain an action in the form
\begin{eqnarray}\label{actp1}
S_{NR}&=&-\frac{\tau_p}{2}\int d^{p+1}\xi V(T)N\sqrt{h}
[-\nabla_nX^a\nabla_nX_a+h^{ij}\partial_iX^a\partial_jX_a]
-\nonumber \\
&-&\frac{\tau_p}{4}\int d^{p+1}\xi
V(T)\sqrt{h}N[h^{ik}h^{jl}(\partial_i A_j-\partial_j A_i)
(\partial_k A_l-\partial_l A_k)-\nonumber \\
&-&2h^{ij}(\mL_n A_i-a_i A_{\bn}-D_iA_{\bn})
(\mL_n A_j-a_j A_{\bn}-D_j A_{\bn})] \ , \nonumber \\
\end{eqnarray}
    where
    \begin{equation}
    \mL_n A_i=\frac{1}{N}(\partial_0 A_i-N^k\partial_k A_i-
    \partial_i N^k A_k) \ , \quad  a_i=\frac{\partial_i N}{N} \ ,
    \end{equation}
and where
\begin{equation}
\nabla_n X^a=\frac{1}{N}(\partial_0 X^a-N^i\partial_i X^a) \ .
\end{equation}
Finally note that $D_i$ is covariant derivative compatible with the metric $h_{ij}$.
Now we are ready to find conjugate momenta from the action
(\ref{actp1}). First of all we obtain momentum conjugate to $A_i$ and $A_{\bn}$
\begin{equation}\label{defpi}
\pi^i=\frac{\delta L}{\delta \partial_0 A_i}=\tau_p
V(T)\sqrt{h}h^{ij}(\mL_n A_j-a_j A_{\bn}-D_j A_{\bn}) \ , \quad
\pi_{\bn}\approx 0  \ .
\end{equation}
Further the momentum conjugate to $X^a$ is equal to
\begin{equation}\label{defpa}
p_a=\frac{\delta L}{\delta \partial_0 X^a}=
\tau_p V(T)\sqrt{h}\nabla_n X_a \ .
\end{equation}
In case of $p_I$ we have to be more careful
since  $N$ and $N_i$ explicitly depend on $\partial_0Y^I$.
Using
\begin{equation}
\frac{\partial N}{\partial_0 Y^I}=\frac{-V_{IJ}\partial_0Y^J}{N} \ , \quad
\frac{\partial N_i}{\partial_0 Y^I}=\partial_i Y_I
\end{equation}
we find after some algebra
\begin{eqnarray}\label{defPi}
\Pi_I&\equiv& p_I+\partial_i Y_I h^{ij}\partial_j X^a p_a+
\partial_k Y_I h^{kl}\partial_k A_i \pi^i-
\partial_l Y_I h^{lk}\partial_i[A_k\pi^i]=
\nonumber \\
&=&-\frac{V_{IJ}\partial_0Y^J}{N}
\left[\frac{1}{2\tau_p V\sqrt{h}}p_a p^a +\frac{\tau_p}{2}V
\sqrt{h}h^{ij}\partial_i X^a\partial_j X_a+\right.\nonumber \\
&+&\left.\frac{1}{2\tau_pV\sqrt{h}}
\pi^i h_{ij}\pi^j+\frac{\tau_p}{4}V\sqrt{h}
h^{ij}h^{kl}F_{ij}F_{kl}+A_{\bn}\partial_i \pi^i \right] \ , \quad
F_{ij}=\partial_i A_j-\partial_j A_i \ .
\nonumber \\
\end{eqnarray}
To proceed  further it is important to stress that $V_{IJ}$
obey the relations
\begin{eqnarray}\label{VIJrel}
\partial_i Y^I V_{IJ}=0 \ , \quad
V_{IJ}\eta^{JK}V_{KL}=V_{IJ}V^J_{ \ L}=V_{IL} \ . \nonumber \\
\end{eqnarray}
Using the first  formula in (\ref{defPi}) we obtain
following primary constraint
\begin{eqnarray}
\mH_i=\partial_i Y^I\Pi_I=
p_I\partial_i Y^I+\partial_i X^a p_a+
\partial_i A_k \pi^k-
\partial_k[A_i\pi^k]\approx 0 \ .
\nonumber \\
\end{eqnarray}
On the other hand if we calculate $\Pi_I V^{IJ}\Pi_J$ defined
in (\ref{defPi}) and use (\ref{VIJrel}) we find
\begin{eqnarray}
& &\tmH_T\equiv V^2\Pi_I V^{IJ}\Pi_J+\nonumber \\
&+ &(\frac{1}{2\tau_p \sqrt{h}}p^a p_a +\frac{\tau_p}{2}V^2
\sqrt{h}h^{ij}\partial_i X^a\partial_j X_a+\frac{1}{2\tau_p\sqrt{h}}
\pi^i h_{ij}\pi^j+
\frac{\tau_p}{4}V^2\sqrt{h}h^{ij}h^{kl}F_{ij}F_{kl}
+VA_{\bn}\partial_i\pi^i)^2 \nonumber \\
&\equiv &  V^2 \Pi_I \eta^{IJ}\Pi_J+\mK^2\approx 0 \ .
\nonumber \\
\end{eqnarray}
If we compare this constraint with the constraints
derived in case of non-relativistic string theory
\cite{Gomis:2004ht} we find an important difference since
it was argued in \cite{Gomis:2004ht} that the Hamiltonian constraints
in non-relativistic theories should be linear in $p_I$ while in our case the constraint
$\tmH_T$ is quadratic in $p_I$ and  cubic in momenta $p_a$. In order to resolve this issue
note that by definition we have that
 $-\Pi_I \eta^{IJ}\Pi_J$ is positive. Then we can write $\tmH_T$ as
\begin{equation}
\tmH_T=\mK^2-(\sqrt{-V^2\Pi_I\eta^{IJ}\Pi_J})^2=
(\mK-V\sqrt{-\Pi_I \eta^{IJ}\Pi_J})(\mK+V\sqrt{-\Pi_I\eta^{IJ}
    \Pi_J})\approx 0 \ .
\end{equation}
We see that there are two branches that can be considered as constraints.
However
the condition that $\mK+V\sqrt{-\Pi_I V^{IJ}\Pi_J}$ is equal to zero
implies that $\mK$ and $V\sqrt{-\Pi_I V^{IJ}\Pi_J}$ should be equal to
zero separately  since $\mK$ is manifestly positive definite. In other words we would
have over constrained system. For that reason it is natural to consider an expression in the second bracket as a constraint and hence we propose that the Hamiltonian constraint has the form
\begin{equation}
\mH_T=\mK-V\sqrt{-\Pi_IV^{IJ}\Pi_J} \approx 0 \ .
\end{equation}
Finally we determine form of the extended Hamiltonian. Using
(\ref{defpi}),(\ref{defpa}) and (\ref{defPi}) we easily determine
the bare Hamiltonian
\begin{eqnarray}
H_B=\int d^p\xi (p_I \partial_0 Y^I+p_a\partial_0 X^a+\pi^i\partial_0 A_i-\mL)=
\int d^p\xi \pi^i\partial_i A_{\bn} \
\nonumber \\
\end{eqnarray}
so that the extended Hamiltonian with primary constraint included has the form
\begin{equation}\label{defHE}
H_E=\int d^p\xi (\lambda^0\mH_T+\lambda^i\mH_i-\partial_i \pi^iA_{\bn}) \ ,
\end{equation}
where in the following we consider $\mK$  without the term $A_{\bn}\partial_i \pi^i$ which is
proportional to the secondary constraint $\mG=\partial_i \pi^i\approx 0$ that
arises from the requirement of the preservation of the constraint $\pi_{\bn}
\approx 0$. In the next section we determine algebra of constraints.
\section{Algebra of Constraints}\label{fourth}
Now we are ready to determine algebra of constraints. As usual we introduce
smeared form of these constraints
\begin{equation}
\bT_T(f)=\int d^p\xi f \mH_T \ ,\quad \bT_S(f^i)=
\int d^p\xi f^i\mH_i \ ,
\end{equation}
where $f,f^i$ are functions of $\xi^\alpha$.
First of all we have
\begin{eqnarray}
& &\pb{\bT_S(f^i),X^a}=-f^i\partial_i X^a \ , \quad  \pb{\bT_S(f^i),p_a}=
-\partial_i (f^ip_a) \ , \nonumber \\
& &\pb{\bT_S(f^i),Y^I}=-f^i\partial_i Y^I \ , \quad  \pb{\bT_S(f^i),p_I}=
-\partial_i(f^ip_I) \ ,  \nonumber \\
& &\pb{\bT_S(f^i),A_i}=-f^j\partial_jA_i+\partial_i f^jA_j \ , \quad
\pb{\bT_S(f^i),\pi_i}=-\partial_j(f^j \pi^i)+\partial_k(f^i\pi^k)  \
\nonumber \\
\end{eqnarray}
and consequently
\begin{eqnarray}
& &\pb{\bT_S(f^i),F_{ij}}=-f^k\partial_k F_{ij}-\partial_i f^k F_{kj}-
F_{ik}\partial_j f^k \ , \nonumber \\
& &\pb{\bT_S(f^i),h_{ij}}=-f^k\partial_k h_{ij}-\partial_i f^k h_{kj}-
h_{ik}\partial_j f^k \ , \nonumber \\
& &\pb{\bT_S(f^i),h^{ij}}=-f^k\partial_k h^{ij}+\partial_k f^i h^{kj}+
h^{ik}\partial_k f^j \ , \nonumber \\
& &\pb{\bT_S(f^i),\sqrt{h}}=-f^i\partial_i \sqrt{h}-\partial_i f^i\sqrt{h} \ .
 \nonumber \\
\end{eqnarray}
Using these results we easily find
\begin{eqnarray}\label{pbTTS}
& &\pb{\bT_S(f^i),\bT_T(g)}=
\bT_T(f^i\partial_i g) \ ,  \nonumber \\
& &\pb{\bT_S(f^i),\bT_S(g^i)}=\bT_S(f^k\partial_k g^j-g^k\partial_k f^j) \ . \nonumber \\
\end{eqnarray}
Finally we proceed to the calculation of the Poisson bracket
\begin{eqnarray}\label{bTNM}
\pb{\bT_T(f),\bT_T(g)}  \ .
\nonumber \\
\end{eqnarray}
First  of all we observe that we can write $\Pi_I$, using
the constraint $\mH_i$, as
\begin{equation}
\Pi_I=
p_I+\partial_i Y_I
h^{ij}\mH_j-\partial_j Y_I h^{ji}\partial_i Y^Kp_K\approx V_I^K p_K \ .
\end{equation}
Then we obtain very important relation
\begin{eqnarray}
& &\pb{\Pi_I(\xi),h_{ij}(\xi')}=
2V_I^J\partial_i\partial_j Y_J(\xi)\delta(\xi-\xi')  \nonumber \\
\end{eqnarray}
and also
\begin{eqnarray}
\pb{\Pi_I(\xi),h^{ij}(\xi)}=-h^{ik}(\xi')
\pb{\Pi_I(\xi),h_{kl}(\xi')}h^{lj}(\xi')=
-[2h^{ik}h^{jl}V_I^J\partial_k\partial_l Y_J](\xi)\delta(\xi-\xi')
\ . \nonumber \\
\end{eqnarray}
These Poisson brackets are local which implies
\begin{eqnarray}
\pb{\int d^p\xi f\sqrt{-\Pi_I V^{IJ}\Pi_J},\int d^p\xi'  g \mK}+
\pb{\int d^p\xi f\mK,\int d^p\xi'g\sqrt{-\Pi_I V^{IJ}\Pi_J}}=0 \ .
\nonumber \\
\end{eqnarray}
As a result the calculation of the Poisson brackets
(\ref{bTNM}) splits to the calculations of two Poisson brackets.
The first one is equal to
\begin{eqnarray}\label{bT1}
\pb{\int d^p\xi f\mK,\int d^p\xi'g \mK}
=\int d^p\xi (f\partial_i g-g\partial_i f)h^{ij}
V^2(p_a \partial_j X^a+F_{jk}\pi^k)  \ . \nonumber \\
\end{eqnarray}
On the other hand the calculation of the second one
is more involved
\begin{eqnarray}\label{BT2}
& &\pb{\int d^p\xi fV\sqrt{-\Pi_I \eta^{IJ}
    \Pi_J}(\xi),\int d^p\xi'gV\sqrt{-\Pi_I\eta^{IJ}
\Pi_J}(\xi')}=
\nonumber \\
&=&\pb{\int d^p\xi fV\sqrt{-p_I V^{IJ}
        p_J}(\xi),\int d^p\xi'gV\sqrt{-p_IV^{IJ}
        p_J}(\xi')}=
\nonumber \\
&=&\int d^p\xi (f\partial_i g-g\partial_i f)V^2
h^{ij}\partial_j Y^Mp_M \ ,   \nonumber \\
\end{eqnarray}
where
\begin{equation}
V^{IJ}=\eta^{IJ}-\partial_i Y^I h^{ij}\partial_j Y^J \ ,
\end{equation}
and we used the fact that
\begin{eqnarray}& &
\pb{p_I(\xi),V^{KL}(\xi')}=
\partial'_i \delta(\xi-\xi')\delta_I^K h^{ij}(\xi')\partial'_j Y^L(\xi')+
\partial'_iY^K(\xi')h^{ij}(\xi')\partial'_j \delta(\xi-\xi')\delta^L_I+\nonumber \\
&+&\partial'_i Y^K h^{ik}(\xi')
\partial'_j Y^L(\xi')h^{jl}(\xi')(\partial'_k \delta(\xi-\xi')\partial'_l
Y_I(\xi')+\partial'_k Y_I(\xi')\partial'_l\delta(\xi-\xi')) \ , \nonumber \\
\end{eqnarray}
where $\partial'_i\equiv \frac{\partial}{\partial \xi'_i}$. We also
used the fact that $V^{IJ}\partial_i Y_J=0.$ Collecting  (\ref{bT1})
and (\ref{BT2}) together we finally obtain
\begin{eqnarray}
\pb{\bT_T(f),\bT_T(g)}&=&
\int d^p\xi (f\partial_ig-g\partial_if)V^2 h^{ij}
(p_a \partial_j X^a+F_{jk}\pi^k+p_M\partial_j Y^M)=
\nonumber \\
&=&
\bT_S((f\partial_i g-g\partial_if)V^2h^{ij}) \ .
\nonumber \\
\end{eqnarray}
This result shows that $\mH_T$  and $\tmH_i$ are  first class constraints which
 is a consequence of the fact that
non-relativistic limit does not affect world-volume structure of the
theory which is still fully diffeomorphism invariant. In the next
section we will analyze the equations of motion at the tachyon
vacuum.
\section{Equations of Motion}\label{fifth}
Now we determine equations of motion for non-relativistic non-BPS Dp-brane in the canonical formulation. Using the extended form of the Hamiltonian $H_E$
(\ref{defHE})
we find
\begin{eqnarray}
\partial_0 X^\mu&=&\pb{X^\mu,H_E}=
\lambda^0 \frac{VV^{\mu J}\Pi_J}{\sqrt{
-\Pi_IV^{IJ}\Pi_J}}+\lambda^i\partial_i X^\mu \ ,\nonumber \\
\partial_0 p_\mu&=&\pb{p_\mu,H_E}=
\nonumber \\
&=&\int d^p\xi \frac{\lambda^0 V}{\sqrt{-\Pi_IV^{IJ}\Pi_J}}
\pb{p_\mu,\Pi_I(\xi)}V^{IJ}\Pi_J(\xi)+
2\partial_i\left[\lambda^0 \frac{\delta \mK}{\delta h_{ij}}
\partial_j X_\mu\right]+\partial_i(\lambda^ip_\mu) \ , \nonumber \\
\end{eqnarray}
where the Poisson bracket $\pb{p_\mu,\Pi_I}$ is complicated expression
whose explicit form is not important for us. At the same way we determine
the equations of motion for the tachyon
\begin{eqnarray}
\partial_0 T&=&\pb{T,H_E}=
\lambda^0 \frac{VV^{TJ}\Pi_J}{\sqrt{
        -\Pi_IV^{IJ}\Pi_J}}+\lambda^i\partial_i T \ ,\nonumber \\
\partial_0 p_T&=&\pb{p_T,H_E}
=\int d^p\xi \frac{\lambda^0 V}{\sqrt{-\Pi_IV^{IJ}\Pi_J}}
\pb{p_T,\Pi_I(\xi)}V^{IJ}\Pi_J(\xi)+
\frac{dV}{dT}\sqrt{-\Pi_I\Pi^I}+\nonumber \\
&+&
2\partial_i\left[\lambda^0 \frac{\delta \mK}{\delta h_{ij}}
\partial_j T\right]+\partial_i(\lambda^ip_T)
-\lambda^0\frac{\delta \mK}{\delta V}\frac{dV}{dT}
 \ , \nonumber \\
\end{eqnarray}
while the equation of motion for $X^a,p_a$ have simpler form
\begin{eqnarray}
\partial_0 X^a&=&\pb{X^a,H_E}=\lambda^0\frac{p_a}{\tau_p\sqrt{h}}
+\lambda^i\partial_i X^a \ ,  \nonumber \\
\partial_0 p_a&=&\pb{p_a,H_E}=\partial_i\left[\lambda^0V^2\sqrt{h}h^{ij}
\partial_j X_a\right]+\partial_i (\lambda^i p_a)
 \ .  \nonumber \\
\end{eqnarray}
Finally we determine
 equations of motion for $A_i,\pi^i$
\begin{eqnarray}
\partial_0 A_i&=&\pb{A_i,H_E}=\frac{\lambda^0 }{\tau_p\sqrt{h}}
h_{ij}\pi^j+\partial_i A_{\bn}+\lambda^j F_{ji} \ , \nonumber \\
\partial_0 \pi^i&=&\pb{\pi^i,H_E}=\partial_m[\lambda^0 V^2 \tau_p
\sqrt{h}h^{mk}h^{il}F_{kl}]+\partial_j (\lambda^i\pi^i)-\partial_j
(\lambda^i\pi^j)  \ . \nonumber \\
\end{eqnarray}
Note that there are also constraints equations that have to be obeyed
\begin{equation}
\mH_T= 0 \ , \quad  \mH_i= 0 \ , \quad \partial_i\pi^i=0 \ .
\end{equation}

Now we would like to analyze these equations of motion at the tachyon vacuum $T=T_{min}$,
where $V(T_{min})=0 \ , \frac{dV}{dT}(T_{min})=0 \ ,
p_T=0 \ ,   \partial_iT=0$. In this case the equations of motion for $X^\mu,p_\mu$ simplify considerably
\begin{eqnarray}
\partial_0 X^\mu=
\lambda^i\partial_i X^\mu \ , \quad
\partial_0 p_\mu=
2\partial_i\left[\lambda^0 \frac{\delta \mK}{\delta h_{ij}}
\partial_j X_\mu\right]+\partial_i(\lambda^ip_\mu) \ , \nonumber \\
\end{eqnarray}
while the equation of motion for $X^a,p_a,A_i$ and $\pi^i$ have the form
\begin{eqnarray}
\partial_0 X^a&=&\lambda^0\frac{p^a}{\tau_p\sqrt{h}}
+\lambda^i\partial_i X^a \ ,  \quad
\partial_0 p_a=\partial_i (\lambda^i p_a) \ ,\nonumber \\
\partial_0 A_i&=&\frac{\lambda^0 }{\tau_p\sqrt{h}}
h_{ij}\pi^j+\partial_i A_{\bn}+\lambda^j F_{ji} \ , \quad
\partial_0 \pi^i=\partial_j (\lambda^i\pi^i)-\partial_j
(\lambda^i\pi^j)  \  \nonumber \\
\end{eqnarray}
together with the set of the constraints
\begin{eqnarray}\label{consvac}
& &\mH_T(T_{min})=\mK(T=T_{min})=\frac{p_a p^a}{2\tau_p \sqrt{h}}+
\frac{\pi^i h_{ij}\pi^j}{2\tau_p \sqrt{h}}=0
\nonumber \\
& & \mH_i=p_a\partial_i X^a+p_\mu\partial_i X^\mu+F_{ij}\pi^j=0 \ .
\nonumber \\
\end{eqnarray}
From the first equation in (\ref{consvac}) we
 see that $\mK$ is a sum of two positive expressions so that the only possibility
to be equal to zero is to demand that $p_a=\pi^i=0$. Then  the
spatial  diffeomorphism constraints imply that $\mH_i=p_\mu
\partial_i X^\mu=0$ which can be obeyed for non-zero
$\partial_iX^\mu$ on condition that $p_\mu=0$ which is also
consistent with the equation of motion for $p_\mu$. Finally the
equations of motion for $X^\mu$ and $X^a$ have the form
\begin{equation}
\partial_0 X^\mu=\lambda^i\partial_i X^\mu \ , \quad
\partial_0 X^a=\lambda^i\partial_i X^a \ .
\end{equation}
We have to demand that spatial derivatives of $X^\mu$ are non-zero in order to have non-singular matrix $h_{ij}$. For that reason we choose following ansatz
\begin{equation}
X^i=\sigma^i \ , i=1,\dots,p-1
\end{equation}
so that the equation of motion for $X^i$ implies $\lambda^i=0 \ ,i=1,\dots,p-1$. On the other hand the equation of motion for $X^0$ has the form
\begin{equation}
\partial_0 X^0=\lambda^p \partial_p X^0
\end{equation}
that can be solved as $X^0=f(\tau+\sigma)$, where $f$ is an arbitrary function and we have imposed the condition that  $\lambda^p=1$. Note that we cannot impose the static gauge condition $X^0=\tau$ since in this case we would find that $h_{ij}$ is singular.
 In the same way we
find that $X^a=v^a(\sigma+\tau)$, where $v^a$ are arbitrary functions.

In summary we find that the equations of motion for non-relativistic non-BPS Dp-brane
at the tachyon vacuum have solutions with rather non physical properties where the dynamics of the transverse coordinates is not related to the conjugate momenta.
In other words it is not possible to identify the tachyon vacuum as the gas of the fundamental non-relativistic strings which is in sharp contrast  with  a similar analysis performed in case of relativistic non-BPS Dp-brane action
\cite{Sen:2000kd,Sen:2003bc,Kluson:2016tgp}. In order to see this in more details let us now review   non-relativistic limit for fundamental string.
\section{Non-Relativistic Limit of Fundamental String and Its Hamiltonian Form}
\label{sixth}
In this section we find non-relativistic string in flat space-time, following
\cite{Batlle:2016iel}. We start with the Nambu-Goto form of string action
\begin{equation}
S=-\ttau_F\int d\tau d\sigma\sqrt{-\det G } \ , \quad
G_{\alpha\beta}=\eta_{MN}\partial_\alpha \tZ^M\partial_\beta \tZ^N \ , \quad
\alpha,\beta=\tau,\sigma
\end{equation}
and perform non-relativistic limit in the form
\begin{equation}
\tZ^\mu=Z^\mu \ , \quad
 \mu=0,1 \ , \quad \tZ^i=\lambda Z^i  \ , \quad  i=2,\dots,d-1 \ ,\quad
\ttau_F=\lambda^{-2}\tau_F \
\end{equation}
so that we can write
\begin{equation}
G_{\alpha\beta}=\partial_\alpha Z^\mu \partial_\beta Z_\mu+
\lambda^2\partial_\alpha Z^i\partial_\beta Z_i\equiv \ba_{\alpha\beta}+
\lambda^2\partial_\alpha Z^i\partial_\beta Z_i
 \
\end{equation}
and hence
\begin{eqnarray}\label{SNRact}
S_{NR}=-\frac{\tau_F}{\lambda^2}
\int d\tau d\sigma \sqrt{-\det \ba}-
\frac{\tau_F}{2}\int d\tau d\sigma \sqrt{-\det \ba}\ba^{\alpha\beta}
\partial_\beta Z^i\partial_\alpha Z_i \ . \nonumber \\
\end{eqnarray}
The first term in (\ref{SNRact})   is proportional to the integral from
total derivative since
\begin{eqnarray}
&-&\det \ba=
[\partial_\tau Z^0\partial_\sigma Z^1-\partial_\tau Z^1\partial_\sigma Z^0]^2=
\nonumber \\
&=&[\partial_\tau (Z^0\partial_\sigma Z^1)-
\partial_\sigma (Z^0\partial_\tau Z^1)]^2 \nonumber \\
\end{eqnarray}
and hence
\begin{equation}
\int d\tau d\sigma \sqrt{-\det \ba}=
\int d\tau d\sigma
[\partial_\tau (Z^0\partial_\sigma Z^1)-
\partial_\sigma (Z^0\partial_\tau Z^1)]
\end{equation}
so that the first term in (\ref{SNRact}) can be ignored. We see that in case
of two world-sheet dimensions the "stringy" non-relativistic limit is well defined even with the absence
of the two form background field and the resulting action has the form
\begin{equation}\label{SNRstring}
S_{NR}=-\frac{\tau_F}{2}\int d\tau d\sigma \sqrt{-\det \ba}
\ba^{\alpha\beta} \partial_\beta Z^i\partial_\alpha Z_i
\ .
\end{equation}
In order to find Hamiltonian form of this action we use $1+1$ notation
as in section (\ref{third}). Explicitly, we have
\begin{eqnarray}
\ba_{\alpha\beta}=\left(\begin{array}{cc}
-n^2+n_\sigma h^{\sigma\sigma}n_\sigma & n_\sigma \\
n_\sigma & h_{\sigma\sigma} \\ \end{array}\right) \ , \quad
\ba^{\alpha\beta}=
\left(\begin{array}{cc}
-\frac{1}{n^2} & \frac{n^\sigma}{n^2} \\
\frac{n^\sigma}{n^2} & h^{\sigma\sigma}-\frac{n^\sigma n^\sigma}{n^2} \\ \end{array}\right)
\end{eqnarray}
so that
\begin{eqnarray}
h_{\sigma\sigma}=\partial_\sigma Z^\mu\partial_\sigma Z_\mu  \ , \quad
h^{\sigma\sigma}=\frac{1}{\partial_\sigma Z^\mu\partial_\sigma Z_\mu} \ , \quad
n_\sigma=\partial_\tau Z^\mu\partial_\sigma Z_\mu \ , \nonumber \\
n^2=-\partial_\tau Z^\mu (\eta_{\mu\nu}-\partial_\sigma Z_\mu
h^{\sigma\sigma}\partial_\sigma Z_\nu)\partial_\tau Z^\nu\equiv -\partial_\tau
Z^\mu v_{\mu\nu}\partial_\tau Z^\nu \ . \nonumber \\
\end{eqnarray}
Using this notation the action (\ref{SNRstring}) has the form
\begin{eqnarray}\label{SNRstring11}
S_{NR}&=&\frac{\tau_F}{2}\int d\tau d\sigma \sqrt{h_{\sigma\sigma}}
(\nabla_n Z^i\nabla_n Z_i-h^{\sigma\sigma}\partial_\sigma Z^i\partial_\sigma Z_i) \ ,  \nonumber \\
\nabla_n Z^i&=&\frac{1}{n}(\partial_\tau Z^i-n^\sigma \partial_\sigma Z^i)  \ .
\end{eqnarray}
Now we are ready to proceed to the Hamiltonian formalism. From
(\ref{SNRstring11}) we find
 following conjugate momenta
\begin{eqnarray}
k_i&=&\tau_F \sqrt{h_{\sigma\sigma}}n\nabla_n Z_i \ , \nonumber \\
k_\mu
&=&\frac{\tau_F}{2}\frac{v_{\mu\nu}\partial_\tau Z^\nu}{n}[\nabla_nZ^i\nabla_n Z_i+h^{\sigma\sigma}\partial_\sigma Z^i\partial_\sigma Z_i]-\tau_F\sqrt{h_{\sigma\sigma}}\partial_
\sigma Z_\mu h^{\sigma\sigma}\partial_\sigma Z^i
\nabla_n Z_i \  \nonumber \\
\end{eqnarray}
that again implies
\begin{eqnarray}\label{pipmustring}
\omega_\mu\equiv k_\mu+\partial_\sigma Z_\mu h^{\sigma\sigma}\partial_\sigma Z^i
k_i=\frac{v_{\mu\nu}\partial_\tau Z^\nu}{n}
\left[\frac{k_i k^i}{2\tau_F\sqrt{h_{\sigma\sigma}}}+\frac{\tau_F}{2}\sqrt{h_{\sigma\sigma}}
h^{\sigma\sigma}\partial_\sigma Z^i\partial_\sigma Z_i\right] \ .
\nonumber \\
\end{eqnarray}
Since $v_{\mu\nu}$ obey the relations
\begin{equation}
v_{\mu\nu}\partial_\sigma Z^\nu=0 \ , \quad  v_{\mu\nu}\eta^{\nu\rho}
v_{\rho\sigma}=v_{\mu\nu} \
\end{equation}
we obtain two primary constraints
\begin{eqnarray}
\mH_\sigma&=&\omega_\mu\partial_\sigma Z^\mu=
k_\mu \partial_\sigma Z^\mu+k_i\partial_\sigma Z^i\approx 0 \ , \nonumber \\
\tmH_\tau&=&\omega_\mu v^{\mu\nu} \omega_\nu+
\left[\frac{k_i k^i}{2\tau_F\sqrt{h_{\sigma\sigma}}}
+\frac{\tau_F}{2}\sqrt{h_{\sigma\sigma}}h^{\sigma\sigma}\partial_\sigma Z^i\partial_\sigma Z_i\right]^2\equiv
\Sigma^2+\omega_\mu v^{\mu\nu} \omega_\nu \approx 0 \ . \nonumber \\
\end{eqnarray}
Since $-\omega_\mu v^{\mu\nu}\omega_\nu$ is positive we can write $\tmH_\tau$
as
\begin{equation}
\tmH_\tau=(\Sigma-\sqrt{-\omega_\mu v^{\mu\nu}\omega_\nu})(
\Sigma+\sqrt{-\omega_\mu v^{\mu\nu}\omega_\nu})  \approx 0 \ .
\end{equation}
Using the same arguments as in section (\ref{third}) we can argue that the
Hamiltonian constraint $\mH_\tau$ has the form
\begin{equation}
\mH_\tau=\Sigma-\sqrt{-\omega_\mu v^{\mu\nu}\omega_\nu}\approx 0
\end{equation}
and we can again show that it is the first class constraints together with
$\mH_\sigma$. Further, it is also easy to see, using
(\ref{pipmustring}), that the bare Hamiltonian is equal to zero so that
the extended Hamiltonian is a sum of two class constraints
\begin{equation}\label{Hamstring}
H_E=\int d\sigma (\lambda^\tau \mH_\tau+\lambda^\sigma \mH_\sigma) \ .
\end{equation}
The equations of motion for $Z^\mu,k_\mu$ that follow from (\ref{Hamstring})
have the form
\begin{eqnarray}
\partial_\tau Z^\mu&=&\pb{Z^\mu,H_E}=
\lambda^\tau \frac{v^{\mu\nu}\omega_\nu}
{\sqrt{-\omega_\mu v^{\mu\nu}\omega_\nu}}
+\lambda^\sigma \partial_\sigma Z^\mu \ , \nonumber \\
\partial_\tau k_\mu&=&\pb{k_\mu,H_E}=
-\partial_\sigma\left[\lambda^\tau \frac{h^{\sigma\sigma}k_\mu \partial_\sigma Z^\nu
k_\nu}{\sqrt{-\omega_\mu v^{\mu\nu}\omega_\nu}}\right]-\nonumber \\
&-&\partial_\sigma\left[\lambda^\tau \frac{\partial_\sigma Z^\rho k_\rho
\partial_\sigma Z^\sigma k_\sigma \partial_\sigma Z_\mu }{h_{\sigma\sigma}^2
\sqrt{-\omega_\mu v^{\mu\nu}\omega_\nu}}\right]
-\partial_\sigma\left[\lambda^\tau \frac{1}{2h_{\sigma\sigma}}\partial_\sigma Z_\mu\Sigma \right]+\partial_\sigma (\lambda^\sigma k_\mu) \ . \nonumber \\
\end{eqnarray}
Finally the equations of motion for $Z^i,p_i$ have the form
\begin{eqnarray}\label{eqZi}
\partial_\tau Z^i&=&\pb{Z^i,H_E}=\lambda^\tau\frac{k^i}{\tau_F \sqrt{h_{\sigma\sigma}}}+
\lambda^\sigma \partial_\sigma Z^i \ , \nonumber \\
\partial_\tau k_i&=&\pb{k_a,H_E}=\partial_\sigma\left[\lambda^\tau
\frac{\tau_F}{\sqrt{h_{\sigma\sigma}}}\partial_\sigma Z_i\right]
+\partial_\sigma (\lambda^\sigma k_i) \ .
\nonumber \\
\end{eqnarray}
Note also that the system has to obey
 the constraints $\mH_\tau=0 \ , \mH_\sigma=0$.
Let us try to solve these equations of motion at the static gauge where
\begin{equation}
Z^1=\sigma \ , \quad  Z^0=\tau
\end{equation}
In this gauge we have $h_{\sigma\sigma}=1, v^{\tau\tau}=-1 \ , v^{\tau\sigma}=0 \ , v^{\sigma\sigma}=0$. Then the equation of motion for $Z^1$ implies $\lambda^\sigma=0$ while the equation of motion for $Z^0$ implies
\begin{equation}
\lambda^\tau=-1 \ .
\end{equation}
On the other hand $k_1$  can be determined using the spatial diffeomorphism constraint $\mH_\sigma=0$ while $k_0$ can be determined using $\mH_\tau=0$
\begin{equation}
k_1=-k_i\partial_\sigma Z^i \  , \quad  k_0=\Sigma \  .
\end{equation}
Using these results in (\ref{eqZi}) we obtain that they simplify
considerably
\begin{equation}
\partial_\tau Z^i=\frac{k^i}{\tau_F} \ , \quad
\partial_\tau K_i=\tau_F \partial_\sigma^2Z_i
\end{equation}
that in the end lead to the wave equation
\begin{equation}
\partial_\tau^2 Z^i-\partial^2_\sigma Z^i=0 \
\end{equation}
that correspond to the string vibrations  in the transverse space with agreement
with \cite{Gomis:2004ht}.

Finally we compare the Hamiltonian of the non-relativistic string
derived here with the Hamiltonian found recently in
\cite{Batlle:2016iel}. In order to do this we introduce light-cone
coordinates
\begin{eqnarray}\label{lcvar}
r&=&Z^0-Z^1 \ , \quad  s=Z^0+Z^1  \ , \nonumber \\
p_r&=&\frac{1}{2}(p_0-p_1) \ , \quad  p_s=\frac{1}{2}(p_0+p_1) \ ,
\nonumber \\
\end{eqnarray}
so that
$\pb{r,p_r}=1 \ , \pb{r,p_s}=0 \ , \pb{s,p_s}=1 \ ,
\pb{s,p_r}=0 $.
In terms of the variables (\ref{lcvar}) we find explicit form
of $v^{\mu\nu}$ and $h_{\sigma\sigma}$
\begin{eqnarray}
  h_{\sigma\sigma}&=&-r's'\ , \quad  r'\equiv \partial_\sigma r \ ,  \quad
\dot{r}=\partial_\tau r \ ,
\nonumber \\
v^{\tau\tau}&=&-1+\frac{1}{4r's'}(r'^2+2r's'+s'^2) \ , \quad
v^{\tau\sigma}=\frac{1}{4r's'}(s'^2-r'^2) \ , \nonumber \\
v^{\sigma\tau}&=&\frac{1}{4r's'}(s'^2-r'^2) \ , \quad  v^{\sigma\sigma}=1+\frac{1}{4r's'}
(r'^2-2r's'+s'^2) \ . \nonumber \\
\end{eqnarray}
Using these formulas we easily determine Hamiltonian constraint in the
light cone variables to be equal to
\begin{equation}
\mH_\tau=\frac{1}{\sqrt{-r's'}}
[p_ip^i+\tau_F^2 Z'_iZ'_i]-
\frac{1}{\sqrt{-r's'}}(p_r r'-p_s s')\approx 0
\end{equation}
that agrees with the Hamiltonian constraint found in
\cite{Batlle:2016iel}. For reader's convenience we repeat the
Hamiltonian analysis performed in  \cite{Batlle:2016iel}
in  Appendix.
\begin{appendix}
\section{Hamiltonian Analysis of Non-Relativistic Fundamental String
    in Light-Cone Formulation}\label{A}
In this Appendix we review the Hamiltonian analysis of non-relativistic
string in the light cone variables. We define these variables as in
previous section
\begin{equation}
Z^0=\frac{1}{2}(r+s) \ , \quad Z^1=\frac{1}{2}(s-r)
\end{equation}
so that we easily find that matrix $\ba_{\alpha\beta}$ and its
inverse $\ba^{\alpha\beta}$ have the form
\begin{eqnarray}
& &\ba_{\tau\tau}=-\dot{r}\dot{s} \ ,  \quad
\ba_{\tau\sigma}=-\frac{1}{2}(\dot{r}s'+\dot{s}r') \ , \quad
\ba_{\sigma\sigma}=-r's'
\ , \nonumber \\
& &\ba^{\tau\tau}
=\frac{4r's'}{(\dot{r}s'-
    \dot{s}r')^2} \ ,
\quad
\ba^{\tau\sigma}=
-2\frac{(\dot{r}s'+\dot{s}r')}{(\dot{r}s'-\dot{s}r')^2} \ ,
\quad
\ba^{\sigma\sigma}=
\frac{4\dot{r}\dot{s}}{(\dot{r}s'-
    \dot{s}r')^2}
 \  \nonumber \\
\end{eqnarray}
so that non-relativistic string action has the form
\begin{eqnarray}\label{actlc}
S^{l.c.}_{NR}=
\tau_F\int d\tau d\sigma \frac{1}{\dot{r}s'-\dot{s}r'}
(-r's'\dot{Z}^i\dot{Z_i}+(\dot{r}s'+\dot{s}r')\dot{Z}^i Z'_i-\dot{r}\dot{s}
Z'^iZ'_i)
\equiv \int d\tau d\sigma \mL_{NR} \ .
\nonumber \\
\end{eqnarray}
From (\ref{actlc}) we find conjugate momenta
\begin{eqnarray}\label{momlightcone}
p_r&=&\frac{\partial \mL_{NR}}{\partial \dot{r}}=
-\frac{s'}{(\dot{r}s'-\dot{s}r')}\mL_{NR}+\frac{\tau_{F}}{\dot{r}s'-\dot{s}r'}
(s'\dot{Z}^iZ'_i-\dot{s}Z'^iZ'_i) \ , \nonumber \\
p_s&=&\frac{\partial \mL_{NR}}{\partial \dot{s}}=
\frac{r'}{(\dot{r}s'-\dot{s}r')}\mL_{NR}+\frac{\tau_{F}}{\dot{r}s'-\dot{s}r'}(r'
\dot{Z}^iZ'_i-\dot{r}Z'^iZ'_i) \ ,\nonumber \\
\nonumber \\
p_i&=&\frac{\partial \mL_{NR}}{\partial \dot{Z}^i}=
\frac{\tau_{F}}{\dot{r}s'-\dot{s}r'}
(-2r's'\dot{Z}_i+(\dot{r}s'+\dot{s}r')Z'_i) \  \nonumber \\
\end{eqnarray}
so that we easily find that the bare Hamiltonian vanish
\begin{equation}
H_B=\int d\sigma (p_r \dot{r}+p_s\dot{s}+p_i\dot{Z}^i-\mL_{NR})=0 \ .
\end{equation}
On the other hand using (\ref{momlightcone}) we find following primary constraints
\begin{equation}
\mH_\sigma=p_r r'+p_s s'+p_i Z'^i\approx 0 \
\end{equation}
and also
\begin{eqnarray}\label{lightconeconstr}
\mH^{lc}_\tau\equiv -p_r r'+p_s s'+ \frac{1}{2\tau_F}\left(p_ip^i+\tau_F^2Z'_iZ'^i\right)\approx 0
\nonumber \\
\end{eqnarray}
that were previously derived in \cite{Batlle:2016iel}.
\end{appendix}

\acknowledgments{This  work  was
    supported by the Grant Agency of the Czech Republic under the grant
    P201/12/G028. }


\begin{thebibliography}{20}

\bibitem{Maldacena:1997re}
J.~M.~Maldacena,
\emph{``The Large N limit of
     superconformal field theories and supergravity,''}
Int.\ J.\ Theor.\ Phys.\  {\bf 38} (1999) 1113
[Adv.\ Theor.\ Math.\ Phys.\  {\bf 2} (1998) 231]
doi:10.1023/A:1026654312961
[hep-th/9711200].

\bibitem{Gubser:1998bc}
S.~S.~Gubser, I.~R.~Klebanov and A.~M.~Polyakov,
\emph{``Gauge theory correlators
    from noncritical string theory,''}
Phys.\ Lett.\ B {\bf 428} (1998) 105
doi:10.1016/S0370-2693(98)00377-3
[hep-th/9802109].

\bibitem{Witten:1998qj}
E.~Witten,
\emph{``Anti-de Sitter space and holography,''}
Adv.\ Theor.\ Math.\ Phys.\  {\bf 2} (1998) 253
[hep-th/9802150].



\bibitem{Hartnoll:2016apf}
S.~A.~Hartnoll, A.~Lucas and S.~Sachdev,
\emph{``Holographic quantum matter,''}
arXiv:1612.07324 [hep-th].

\bibitem{Horava:2009uw}
P.~Horava,
\emph{``Quantum Gravity at a Lifshitz Point,''}
Phys.\ Rev.\ D {\bf 79} (2009) 084008
doi:10.1103/PhysRevD.79.084008
[arXiv:0901.3775 [hep-th]].

\bibitem{Wang:2017brl}
A.~Wang,
\emph{``Ho\v{r}ava Gravity
     at a Lifshitz Point: A Progress Report,''}
arXiv:1701.06087 [gr-qc].





\bibitem{Bergshoeff:2017btm}
E.~Bergshoeff, J.~Gomis, B.~Rollier, J.~Rosseel and T.~ter Veldhuis,
\emph{``Carroll versus Galilei Gravity,''}
arXiv:1701.06156 [hep-th].


\bibitem{Bergshoeff:2016lwr}
E.~A.~Bergshoeff and J.~Rosseel,
\emph{``Three-Dimensional Extended Bargmann Supergravity,''}
Phys.\ Rev.\ Lett.\  {\bf 116} (2016) no.25,  251601
doi:10.1103/PhysRevLett.116.251601
[arXiv:1604.08042 [hep-th]].

\bibitem{Hartong:2015xda}
J.~Hartong,
\emph{``Gauging the Carroll Algebra
    and Ultra-Relativistic Gravity,''}
JHEP {\bf 1508} (2015) 069
doi:10.1007/JHEP08(2015)069
[arXiv:1505.05011 [hep-th]].

\bibitem{Bergshoeff:2015uaa}
E.~Bergshoeff, J.~Rosseel and T.~Zojer,
\emph{``Newton–Cartan (super)gravity as
     a non-relativistic limit,''}
Class.\ Quant.\ Grav.\  {\bf 32} (2015) no.20,  205003
doi:10.1088/0264-9381/32/20/205003
[arXiv:1505.02095 [hep-th]].


\bibitem{Andringa:2010it}
R.~Andringa, E.~Bergshoeff, S.~Panda and M.~de Roo,
\emph{``Newtonian Gravity and the Bargmann Algebra,''}
Class.\ Quant.\ Grav.\  {\bf 28} (2011) 105011
doi:10.1088/0264-9381/28/10/105011
[arXiv:1011.1145 [hep-th]].


\bibitem{Bergshoeff:2015ija}
E.~Bergshoeff, J.~Rosseel and T.~Zojer,
\emph{``Newton-Cartan supergravity with torsion and Schrödinger supergravity,''}
JHEP {\bf 1511} (2015) 180
doi:10.1007/JHEP11(2015)180
[arXiv:1509.04527 [hep-th]].










\bibitem{Gomis:2000bd}
J.~Gomis and H.~Ooguri,
\emph{``Nonrelativistic closed string theory,''}
J.\ Math.\ Phys.\  {\bf 42} (2001) 3127
doi:10.1063/1.1372697
[hep-th/0009181].


\bibitem{Danielsson:2000gi}
U.~H.~Danielsson, A.~Guijosa and M.~Kruczenski,
\emph{``IIA/B, wound and wrapped,''}
JHEP {\bf 0010} (2000) 020
doi:10.1088/1126-6708/2000/10/020
[hep-th/0009182].


\bibitem{Gomis:2004pw}
J.~Gomis, K.~Kamimura and P.~K.~Townsend,
\emph{``Non-relativistic superbranes,''}
JHEP {\bf 0411} (2004) 051
doi:10.1088/1126-6708/2004/11/051
[hep-th/0409219].

\bibitem{Gomis:2005pg}
J.~Gomis, J.~Gomis and K.~Kamimura,
\emph{``Non-relativistic superstrings:
     A New soluble sector of AdS(5) x S**5,''}
JHEP {\bf 0512} (2005) 024
doi:10.1088/1126-6708/2005/12/024
[hep-th/0507036].

\bibitem{Kluson:2006xi}
J.~Kluson,
\emph{``Non-Relativistic Non-BPS Dp-brane,''}
Nucl.\ Phys.\ B {\bf 765} (2007) 185
doi:10.1016/j.nuclphysb.2006.12.010
[hep-th/0610073].


\bibitem{Batlle:2016iel}
C.~Batlle, J.~Gomis and D.~Not,
\emph{``Extended Galilean symmetries of non-relativistic strings,''}
JHEP {\bf 1702} (2017) 049
doi:10.1007/JHEP02(2017)049
[arXiv:1611.00026 [hep-th]].








\bibitem{Gomis:2016zur}
J.~Gomis and P.~K.~Townsend,
\emph{``The Galilean Superstring,''}
arXiv:1612.02759 [hep-th].






%
%



\bibitem{Kluson:2017fam}
J.~Kluson,
\emph{``Carroll Limit of Non-BPS Dp-Brane,''}
arXiv:1702.08685 [hep-th].




    \bibitem{Bergshoeff:2014jla}
    E.~Bergshoeff, J.~Gomis and G.~Longhi,
\emph{``Dynamics of Carroll Particles,''}
    Class.\ Quant.\ Grav.\  {\bf 31} (2014) no.20,  205009
    doi:10.1088/0264-9381/31/20/205009
    [arXiv:1405.2264 [hep-th]].

\bibitem{Bergshoeff:2015wma}
E.~Bergshoeff, J.~Gomis and L.~Parra,
\emph{``The Symmetries of the Carroll Superparticle,''}
J.\ Phys.\ A {\bf 49} (2016) no.18,  185402
doi:10.1088/1751-8113/49/18/185402
[arXiv:1503.06083 [hep-th]].


\bibitem{Bergshoeff:2014uea}
E.~A.~Bergshoeff, J.~Hartong and J.~Rosseel,
\emph{``Torsional Newton–Cartan geometry and the Schrödinger algebra,''}
Class.\ Quant.\ Grav.\  {\bf 32} (2015) no.13,  135017
doi:10.1088/0264-9381/32/13/135017
[arXiv:1409.5555 [hep-th]].



\bibitem{Cardona:2016ytk}
B.~Cardona, J.~Gomis and J.~M.~Pons,
\emph{``Dynamics of Carroll Strings,''}
JHEP {\bf 1607} (2016) 050
doi:10.1007/JHEP07(2016)050
[arXiv:1605.05483 [hep-th]].



\bibitem{Sen:1999md}
A.~Sen,
\emph{``Supersymmetric world volume action for nonBPS D-branes,''}
JHEP {\bf 9910} (1999) 008
doi:10.1088/1126-6708/1999/10/008
[hep-th/9909062].


\bibitem{Bergshoeff:2000dq}
E.~A.~Bergshoeff, M.~de Roo, T.~C.~de Wit, E.~Eyras and S.~Panda,
\emph{``T duality and actions for nonBPS D-branes,''}
JHEP {\bf 0005} (2000) 009
doi:10.1088/1126-6708/2000/05/009
[hep-th/0003221].



\bibitem{Kluson:2000iy}
J.~Kluson,
\emph{``Proposal for nonBPS D-brane action,''}
Phys.\ Rev.\ D {\bf 62} (2000) 126003
doi:10.1103/PhysRevD.62.126003
[hep-th/0004106].



\bibitem{Sen:2004nf}
A.~Sen,
\emph{``Tachyon dynamics in open string theory,''}
Int.\ J.\ Mod.\ Phys.\ A {\bf 20} (2005) 5513
doi:10.1142/S0217751X0502519X
[hep-th/0410103].


\bibitem{Sen:2000kd}
A.~Sen,
\emph{``Fundamental strings in open string theory at the tachyonic vacuum,''}
J.\ Math.\ Phys.\  {\bf 42} (2001) 2844
doi:10.1063/1.1377037
[hep-th/0010240].

\bibitem{Sen:2003bc}
A.~Sen,
\emph{``Open and closed strings from unstable D-branes,''}
Phys.\ Rev.\ D {\bf 68} (2003) 106003
doi:10.1103/PhysRevD.68.106003
[hep-th/0305011].



\bibitem{Kluson:2016tgp}
J.~Kluson,
\emph{``Note about unstable D-branes with dynamical tension,''}
Phys.\ Rev.\ D {\bf 94} (2016) no.4,  046004
doi:10.1103/PhysRevD.94.046004
[arXiv:1605.09510 [hep-th]].

\bibitem{Kluson:2005dr}
J.~Kluson,
\emph{``Remark about non-BPS Dp-brane at the tachyon vacuum moving in curved background,''}
Phys.\ Rev.\ D {\bf 72} (2005) 106005
doi:10.1103/PhysRevD.72.106005
[hep-th/0504062].



\bibitem{Gomis:2004ht}
J.~Gomis and F.~Passerini,
\emph{``Rotating solutions of non-relativistic string theory,''}
Phys.\ Lett.\ B {\bf 617} (2005) 182
doi:10.1016/j.physletb.2005.04.061
[hep-th/0411195].





\bibitem{Erkal:2009xq}
D.~Erkal, D.~Kutasov and O.~Lunin,
\emph{``Brane-Antibrane Dynamics From the Tachyon DBI Action,''}
arXiv:0901.4368 [hep-th].


\bibitem{Kutasov:2004ct}
D.~Kutasov,
\emph{``A Geometric interpretation of the open string tachyon,''}
hep-th/0408073.

\bibitem{Kutasov:2003er}
D.~Kutasov and V.~Niarchos,
\emph{``Tachyon effective actions in open string theory,''}
Nucl.\ Phys.\ B {\bf 666} (2003) 56
doi:10.1016/S0550-3213(03)00498-X
[hep-th/0304045].

\bibitem{Sen:2007cz}
A.~Sen,
\emph{``Geometric tachyon to universal open string tachyon,''}
JHEP {\bf 0705} (2007) 035
doi:10.1088/1126-6708/2007/05/035
[hep-th/0703157 [HEP-TH]].



\bibitem{Gourgoulhon:2007ue}
E.~Gourgoulhon,
\emph{``3+1 formalism and bases of numerical relativity,''}
gr-qc/0703035 [GR-QC].


\bibitem{Arnowitt:1962hi}
R.~L.~Arnowitt, S.~Deser, C.~W.~Misner,
\emph{``The Dynamics of general
    relativity,''}
[gr-qc/0405109].
\end{thebibliography}
\end{document}